\documentclass [epjST,final]{svjour}

\usepackage[english]{babel}
\usepackage{latexsym}
\usepackage{amsmath}
\usepackage{amssymb}
\usepackage{graphics}
\usepackage{subfigure}
\usepackage{epsfig}
\usepackage{color}
\usepackage{hyperref}

\usepackage{subeqnarray}

\hypersetup{
colorlinks=true,
citecolor=blue,
linkcolor=red,
urlcolor=black
}

\newcommand{\Jnature}{Nature (London)}
\newcommand{\Jnatphys}{Nature Phys.}

\newcommand{\Jscience}{Science}

\newcommand{\Jprl}{Phys. Rev. Lett.}
\newcommand{\Jpr}{Phys. Rev.}
\newcommand{\Jpra}{Phys. Rev. A}
\newcommand{\Jprb}{Phys. Rev. B}

\newcommand{\Jrmp}{Rev. Mod. Phys.}

\newcommand{\Jepl}{Europhys. Lett.}
\newcommand{\Jnjp}{New J. Phys.}

\newcommand{\JApplPhysLett}{Appl. Phys. Lett.}

\newcommand{\JJlowT}{J. Low Temp. Phys.}

\newcommand{\Jjetp}{Sov. Phys. JETP}

\newcommand{\Jphysrep}{Phys. Rep.}

\newcommand{\JjphysA}{J. Phys. A: Math. Theor.}

\renewcommand{\i}{\textrm{i}}
\newcommand{\ie}{{i.e.}}
\newcommand{\eg}{{e.g.}}
\newcommand{\etal}{{\it et al.}}

\newcommand{\be}{\begin{equation}}
\newcommand{\beq}{\begin{eqnarray}}
\newcommand{\ee}{\end{equation}}
\newcommand{\eeq}{\end{eqnarray}}

\newcommand{\ch}{\textrm{cosh}}
\newcommand{\sh}{\textrm{sinh}}
\newcommand{\ftheta}[1]{\left[ #1 \right]_{\oplus}}

\newcommand{\e}{\textrm{e}}

\newcommand{\av}[1]{\overline{#1}}

\newcommand{\Vr}{V_\textrm{\tiny R}}
\newcommand{\sigmar}{\sigma_\textrm{\tiny R}}

\newcommand{\eigenvect}{\phi_\textrm{\tiny \textit E}}
\newcommand{\ray}{r}
\newcommand{\phase}{\theta}

\newcommand{\lyap}{\gamma}
\newcommand{\loc}{L_{\textrm{\tiny loc}}}
\newcommand{\lyapE}[1]{\lyap{(#1)}}
\newcommand{\kE}{k_\textrm{\tiny \textit E}}

\newcommand{\fn}[1]{f_{#1}}
\newcommand{\Cor}[1]{C_{#1}}

\newcommand{\TFCor}[1]{\tilde{C_{#1}}}

\newcommand{\Id}{I_{\textrm{\tiny D}}}
\newcommand{\Escreen}{\mathcal{E}}
\newcommand{\Iscreen}{\mathcal{I}}
\newcommand{\lambdalaser}{\lambda_{\textrm{\tiny 0}}}
\newcommand{\lprop}{l}

\newcommand{\kc}{k_\textrm{c}}

\newcommand{\Erf}[1]{\textrm{erf}\left(#1\right)}
\newcommand{\Erfi}[1]{\textrm{erfi}\left(#1\right)}

\newcommand{\distv}{\mathcal{D}}
\newcommand{\distvi}{\distv_{\textrm{i}}}

\newcommand{\distE}{\mathcal{D}_\textrm{\tiny E}}

\newcommand{\Emaxgas}{E_{\textrm{\tiny at}}}

\newcommand{\largdist}{p_\textrm{\tiny w}}
\newcommand{\centredist}{p_\textrm{\tiny at}}
\newcommand{\kat}{k_\textrm{\tiny at}}

\newcommand{\lyapfit}{\gamma_\textrm{\tiny fit}}

\newcommand{\pmax}{p_{\textrm{\tiny cut}}}
\newcommand{\kmax}{k_{\textrm{\tiny cut}}}

\newcommand{\rap}{\rho}
\newcommand{\kappaNot}{\kappa_\textrm{\tiny 0}}

\newcommand{\Ltot}{L_\textrm{\tiny tot}}

\begin{document}

\title{Tailoring Anderson localization by disorder correlations in 1D speckle potentials}

\author{Marie~Piraud \and Laurent~Sanchez-Palencia}
\institute{
  Laboratoire Charles Fabry,
  Institut d'Optique, CNRS, Univ Paris Sud,\\
  2 avenue Augustin Fresnel,
  F-91127 Palaiseau cedex, France
}

\date{\today}

\abstract{
We study Anderson localization of single particles in continuous, correlated,
one-dimensional disordered potentials.
We show that tailored correlations can completely change the energy-dependence
of the localization length.
By considering two suitable models of disorder,
we explicitly show that disorder correlations can lead to a nonmonotonic behavior
of the localization length versus energy.
Numerical calculations performed within the transfer-matrix approach
and analytical calculations performed within the phase formalism up to order three
show excellent agreement and demonstrate the effect.
We finally show how the nonmonotonic behavior of the localization length
with energy can be observed using expanding ultracold-atom gases.
}

\maketitle

\section{Introduction}

Coherent transport in disordered media shows considerable interest in condensed-matter physics, with applications to normal solids~\cite{lee1985,janssen1998}, superconductors~\cite{degennes1995} and superfluids~\cite{reppy1992}.
Coherent process can lead to the spatial localization of wave functions as a result of a subtle interference effect between multiple scattering paths, which survives disorder averaging.
This effect, known as Anderson localization (AL), was first predicted for electronic matter waves~\cite{anderson1958}.
It was later shown to be a universal phenomenon in wave physics~\cite{john1984}, which permitted
the first evidence of AL of classical waves~\cite{wiersma1997,storzer2006,schwartz2007,lahini2008,hu2008}.
The observation of
AL in ultracold gases in one (1D)~\cite{billy2008,roati2008} and three (3D)~\cite{kondov2011,jendrzejewski2012} dimensions has triggered a renewed interest
on matter wave localization~\cite{piraud2012a,piraud2012c}, and paves the way to further investigation of many open questions~\cite{lsp2010}.

Correlated disorder makes AL fascinatingly rich.
Disorder correlations can
change the localization properties, not only quantitatively but also qualitatively.
For instance, disorder correlations with a finite support in momentum space were shown to
  induce effective mobility edges in 1D disorder~\cite{izrailev1999,lsp2007,lsp2008}, which was used to create materials with alternating localizing and almost transparent frequency windows~\cite{kuhl2000},
  to enhance localization in microwave systems~\cite{kuhl2008},
and to
  propose realization of atomic band-pass filters~\cite{plodzien2011}.
Such correlations are also responsible for algebraic localization
of matter waves with broad energy distributions~\cite{billy2008,lsp2007,piraud2011}.
It was also shown that in certain models of disorder with infinite-range correlations
(\ie\ such that the correlation function does not decay to zero at infinite distance)~\cite{kotani1984,garciagarcia2009}
or dimer-like correlations~\cite{sedrakyan2004,sedrakyan2011},
multiple localization-delocalization transition points can appear.
Interestingly,
tailoring the disorder correlations can lead to a counter-intuitive behavior
of the energy dependence of the localization strength~\cite{piraud2012b}.
It can serve to discriminate quantum versus classical localization of particles,
which is of particular interest for ultracold atoms where disordered potentials can be controlled~\cite{clement2006,shapiro2012}.

The latter effect is the subject of the present paper.
For the sake of clarity, consider matter waves in free space
and subjected to a 1D disordered potential.
The macroscopic behavior of AL is intimately related to the microscopic properties of single scattering
from the asperities of the disorder.
This shows up in the strong dependence of the localization length $\loc(E)$
-- \ie\ the characteristic length scale
of exponential decay of localized wavefunctions -- on the energy $E$ and the Fourier component of the disordered potential $V$ at twice the typical particle wave vector,
$\kE \equiv \sqrt{2mE/\hbar}$:
The 1D localization of a particle is dominated by the interference of quantum paths that
are backscattered twice in the disorder with the forward propagating one.
For weak disorder, the leading term is
$\loc(E)^{-1} \propto \TFCor{2}(2\kE)/E$,
where $\TFCor{2}$ is the structure factor (correlation function of the disordered potential in
Fourier space, see below)~\cite{lifshits1988}.
For most models of disorder (\eg\ for $\delta$-correlated, Gaussian-correlated or usual speckle~\cite{clement2006} disorder), $\TFCor{2}$ is a constant or decreasing function,
so that $\loc(E)$ is an increasing function of the particle energy $E$,
which finds an intuitive interpretation as the higher energy the weaker localization.

Structured disorder correlations can however completely change the picture.
If the disordered potential exhibits structures on a length scale
of the order of $\kE^{-1}$, the scattering might not strictly increase with energy\footnote{This effect is similar to the well-known example of
single scattering from double barriers (see for instance Ref.~\cite{basdevant2005})},
and
$\loc(E)$ can counter-intuitively decrease for increasing $E$~\cite{piraud2012b}.
In this work, we study this effect for particles in continuous
disordered potentials with tailored correlations in one dimension.
We first calculate the localization of the single-particle eigenstates,
using both
  numerical calculations, based on the transfer-matrix approach,
and
  the so-called \textit{phase formalism}, which is well suited
for perturbative expansion in 1D transmission schemes for a matterwave of fixed energy.
Lowest-order analytical calculations reproduce the main physics and allow
design of disorder correlations to realize the desired effect~\cite{piraud2012b}.
We study two alternative possibilities, which require only slight modifications
of existing experimental schemes~\cite{billy2008,clement2008,chen2008,dries2010}.
These calculations however show significant deviations with numerical data,
but we show that they are quantitatively accounted for by next-order calculations.
We finally discuss how to observe the nonmonotonic behavior of the localization length with energy
with expanding ultracold-atom gases,
and explicitly show that standard schemes should be adapted.

\section{Tailoring disorder correlations in speckle potentials}
\begin{figure}[t]
\center
\sidecaption
\includegraphics[width=0.75\textwidth]{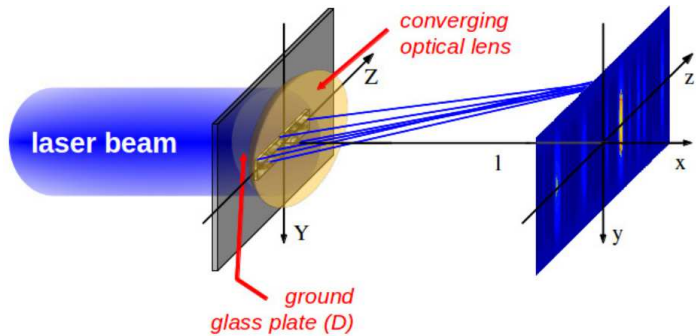}
\caption{
\label{fig:speckle-obtention}
Optical scheme to create a speckle pattern:
A laser beam is diffracted by a ground-glass plate diffuser (D) with a 1D slit
of pupil function $\Id (Z)$, where $Z$ spans the diffuser.
The latter imprints a random phase on the various light paths.
The intensity field, $\Iscreen (z)$, observed in the focal plane of
a converging lens, is a speckle pattern,
which creates a disordered potential $V(z)$ for the atoms.}
\end{figure}

Disorder realized by optical speckle~\cite{goodman2007,clement2006,shapiro2012} is relatively easy to obtain
and its statistical properties are --to some extent-- controllable,
hence making a good candidate for tailoring correlations.
Such disordered potentials have been used in several experiments to investigate the effect of disorder in ultracold-atom systems~\cite{billy2008,clement2008,chen2008,dries2010,kondov2011,jendrzejewski2012}.
A 1D speckle pattern~\cite{goodman2007} is obtained by shining a coherent light beam onto an elongated aperture with a ground-glass plate diffuser,
and observing the diffraction pattern that is created,
for example in the focal plane of a lens (see Fig.~\ref{fig:speckle-obtention}).
The role of the diffuser is mainly to imprint a random phase on the electric field
at each point of its aperture.
It follows from direct application of
the laws of diffraction,
that the electric field at a point
$z$ on the focal plane, $\Escreen(z)$, is the sum of many complex independent random variables that corresponds to the components originating from every point of the plate and interfering in $z$.
It is then a complex Gaussian random variable. In ultracold atom experiments, the atoms are
sensitive to the light intensity, proportional to $\vert\Escreen(z)\vert^2$.
More precisely the atoms experience a potential that can be written
$V(z) = \Vr \left[|\Escreen(z)|^2/\av{|\Escreen|^2}-1\right]$,
so that $\av{V}=0$ and $\av{V^2}=\Vr^2$~\cite{clement2006,shapiro2012}.
The amplitude of the disorder, $\Vr$, can be
  positive (so called 'blue detuned speckle')
or
  negative ('red detuned speckle'),
depending on the sign of the difference of the atom and laser frequencies.
More generally, all statistical properties of the disordered potential $V(z)$
follow from basic laws of optics. For instance,
one can show that the Fourier transform of the autocorrelation function of the potential is the auto-convolution of the pupil function $\Id$ (intensity pattern exiting the diffuser):
\be
\label{eq:c2-autoconv}
\TFCor{2}(k)\propto \int dZ \, \Id\left(Z - \frac{\lambdalaser \lprop}{4 \pi}k\right) \Id\left(Z+\frac{\lambdalaser \lprop}{4 \pi}k\right),
\ee
where $\lambdalaser$ is the laser wavelength
and $\lprop$ is the focal length~\cite{goodman2007}.
For a thin slit of length $2R$ uniformly lit, which is the usual manner to obtain a speckle,
we have $\Id(Z)=I_0 \Theta(R-|Z|)$, with $\Theta$ the Heaviside function
[$\Theta(x)=1$ if $x>0$ and 0 otherwise], and we find~\footnotemark\footnotetext{We use the conventions $\tilde{f}(\kappa)=\int f(u)e^{-i\kappa u}du$.}
\be
\label{eq:C2-1ouv}
\TFCor{2}(k) = \pi \Vr^2 \sigmar \ftheta{1 - {|k| \sigmar}/{2}},
\ee
where $\ftheta{x}=x\Theta(x)$ and $\sigmar = \lambdalaser \lprop/2 \pi R$
is the disorder correlation length. 
Therefore, the power spectrum of the speckle pattern is strictly decreasing with $k$.
Note that
this model has a cut-off in the Fourier components at $\kc=2 \sigmar^{-1}$.
This is a characteristics of speckle potentials, which in particular
leads to the existence of effective mobility edges,
as discussed in Refs.~\cite{lsp2007,lsp2008,lugan2009,gurevich2009}.

Modifying the pupil function $\Id(Z)$ (\ie\ changing the aperture of the diffusive plate or the spatial profile of the incident beam)
allows us to tailor the disorder correlations~\cite{plodzien2011,piraud2012b}.
In this work, we consider two configurations.
In the first configuration,
we propose to put a mask of width $2r$ at the center of the aperture, creating a double-slit.
When doing so, a gap is created in the pupil function
[$\Id(Z)=I_0\Theta(|Z|-r)\Theta(R-|Z|)]$,
leading, for $\rho=r/R>0$, to an increase of the integral~(\ref{eq:c2-autoconv}) on a certain interval of $k$, which is all the more marked that $\rap$ is large:
\be
\label{eq:C2-2ouv}
\TFCor{2}(k) = \frac{\pi \Vr^2 \sigmar}{(1-\rap)^2} \bigg\{
\ftheta{1-\rap-|k|\sigmar}
+ \frac{1}{2} \ftheta{ 1-\rap - \left| |k|\sigmar- {(\rap+1)} \right| } \bigg\}.
\ee
In the second configuration, we propose to illuminate an infinitely-long slit by two mutually coherent Gaussian laser beams of waist $w$ along $Z$ and centered at $Z=\pm \Delta/2$.
The two-point correlation function is then given by
\be
\label{eq:C2-gauss}
\TFCor{2}(k)= \frac{\sqrt{\pi} \Vr^2 \sigmar}{4}
\left[ \e^{-\frac{\left( k\sigmar-\kappaNot \right)^2}{4}} + 2\e^{- \frac{( k\sigmar)^2}{4}} + \e^{- \frac{\left(k\sigmar+\kappaNot \right)^2}{4}} \right],
\ee
with $\sigmar=\lambdalaser\lprop / \pi w$
and
$\kappaNot = 2\Delta/w$.
For $\kappaNot \gtrsim 3.7$, this function also increases on a certain interval of $k$.
We will see in the following that the increase of $\TFCor{2}(k)$
in both the double-slit and the double-Gaussian configurations can lead to an enhancement of localization with energy.

\section{Anderson localization in 1D tailored speckle potentials}
We now study AL of a single particle in these disordered potentials using the so-called phase formalism~\cite{lifshits1988}, which allows for efficient perturbative
expansions~\cite{lugan2009}. Consider a particle of mass $m$ and given energy $E$
in the disordered potential. The corresponding eigenstate $\eigenvect(z)$
is governed by the
1D Schr\"odinger equation
\be
-({\hbar^2}/{2m}) \partial_z^2 \eigenvect(z) + V (z) \eigenvect(z) = E \eigenvect(z).
\label{eq:schro} 
\ee
Then,
write the (real-valued) eigenfunction
$\eigenvect(z) = \ray(z) \sin{[\phase(z)]}$
and its spatial derivative
$\partial_z \eigenvect(z) = \kE \ray(z) \cos{[\phase(z)]}$.
In this representation, Eq.~(\ref{eq:schro}) transforms into the set of equations
\begin{subeqnarray}
\label{eq:schromod}
 \slabel{eq:schromod1}
 \partial_z \phase(z)=\kE \left( 1-\frac{V(z)}{E} \sin^2[\phase(z)] \right)\\
 \slabel{eq:schromod2}
\qquad \ln\left[ \frac{\ray(z)}{\ray(0)} \right]= \kE \int_0^z dz'\frac{V(z')}{2E}\sin[2\phase(z')].
\end{subeqnarray}
Equation~(\ref{eq:schromod1})
can be solved in the form of
a Born-like perturbative series
for the phase $\phase$ in powers of the external potential $V$.
Then, the Lyapunov exponent $\lyapE{E} \equiv \loc(E)^{-1} \equiv \lim_{|z|\rightarrow \infty} \av{\ln[\ray(z)]}/|z|$ can be calculated at each order in $V$ by inserting the result of Eq.~(\ref{eq:schromod1}) into Eq.~(\ref{eq:schromod2}).
It yields\footnote{Note that the first-order term vanishes because $\av{V}=0$.} $\lyapE{E}= \sum_{n \ge 2} \lyap^{(n)}(E)$ with
\be
\label{eq:dvpt-ordres}
\lyap^{(n)} =\frac{1}{\sigmar} \left( \frac{\Vr}{\sqrt{E}} \sqrt{\frac{2m \sigmar^2}{\hbar^2}} \right)^n \fn{n}(\kE \sigmar),
\ee
where
each function $\fn{n}$ depends on the n-point correlation function of the disorder,
$\Cor{n}(z_1,...,z_{n-1})=\av{V(0) V(z_1)...V(z_{n-1})}$~\cite{lugan2009}.
In particular, the leading term of the series
(Born approximation) is
\be
\label{eq:lyap-ordre2}
\fn{2}(\kappa)=\frac{1}{8} \frac{\TFCor{2}(2\kappa/\sigmar)}{\Vr^2\sigmar}.
\ee
This term generally captures most localization properties in 1D disorder,
in particular the effect of the tailored correlations we consider here.
Previous work has however shown some discrepancies between analytic calculations
in the Born approximation and numerical calculations~\cite{piraud2012b}.
Therefore, we will also include the next-order term~\cite{lugan2009}:
\be
\label{eq:lyap-ordre3}
\fn{3}(\kappa)=\frac{-1}{4} \int_{-\infty}^0 du \int_{-\infty}^u dv \frac{\Cor{3}(u\sigmar, v\sigmar)}{\Vr^3} \sin(2\kappa v).
\ee
In addition,
we will perform numerical calculations for the transmission through the 1D disordered potential
of a particle governed by Eq.~(\ref{eq:schro}), using transfer matrix techniques~\cite{mueller2009}.

\begin{figure}[t]
\center
\sidecaption
\includegraphics[width=0.75\textwidth]{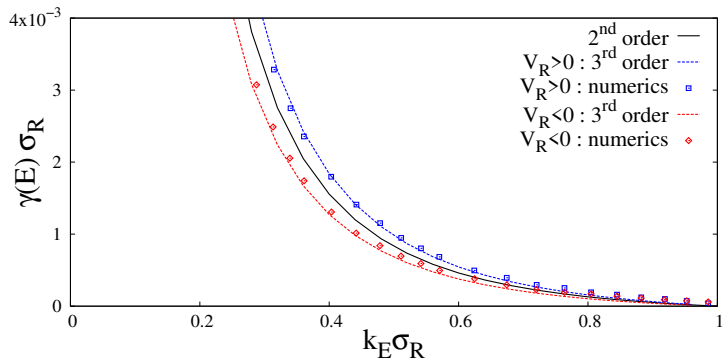}
\caption{
\label{fig-simpleouv}
(Color online)
Lyapunov exponent for both blue- and red-detuned speckles with the standard correlation function~(\ref{eq:C2-1ouv}) and $\Vr=\pm 0.01625(\hbar^2/m \sigmar^2)$.
Shown are
  the numerical results extracted from a transfer matrix method (space step of $\Delta z=0.1 \sigmar$ and total system size of $\Ltot=411775 \sigmar$ with random initial conditions) averaged over 5000 disorder realizations (blue squares: $\Vr>0$, red diamonds: $\Vr<0$), and
  analytical results obtained from the phase formalism up to order $2$ (solid black line) and up to order $3$ (dotted blue and red lines).
}
\end{figure}

Let us start with the standard (single-slit) configuration where the pupil function is uniform and nonzero in the interval $-R<Z<+R$, which corresponds to the correlation function given by Eq.~(\ref{eq:C2-1ouv}).
Figure~\ref{fig-simpleouv} shows the Lyapunov exponent in this configuration,
for both blue- and red-detuned speckle potentials,
with parameters relevant for current experiments~\cite{billy2008}.
The numerical data (blue squares and red diamonds) are averaged over 5000 realizations for each value of $\kE$. The analytic calculations of the Lyapunov exponent in the Born approximation [$\lyap^{(2)}(E)$ given by Eq.~(\ref{eq:dvpt-ordres}) with $n=2$ and Eq.~(\ref{eq:lyap-ordre2}); solid black line], which do not depend on the sign of $\Vr$, fairly reproduce the numerical data. 
In this standard configuration, both numerics and analytic calculation in the Born approximation
confirm that the Lyapunov exponent, \ie\ the localization strength, decreases with increasing particle energy, hence following the intuitive behavior.
We however find a significant discrepancy, which depends on the sign of $\Vr$, between
the numerics and the analytics.
A similar discrepancy was also observed in Ref.~\cite{piraud2012b} but the numerics were not calculating $\lyap(E)$ in a direct way as in the present work.
As is seen on Fig.~\ref{fig-simpleouv}, this discrepancy is very well accounted for by
analytic calculations to the next order in the perturbative series,
$\lyap^{(2)}(E)+\lyap^{(3)}(E)$,
where $\lyap^{(3)}(E)$ is given by Eq.~(\ref{eq:dvpt-ordres}) with $n=3$ and Eq.~(\ref{eq:o3-dblslit}) with $\rho=0$. Note that $\lyap^{(3)}$ depends on the sign of $\Vr$.

\begin{figure}[t]
\center
\sidecaption
\includegraphics[width=0.75\textwidth]{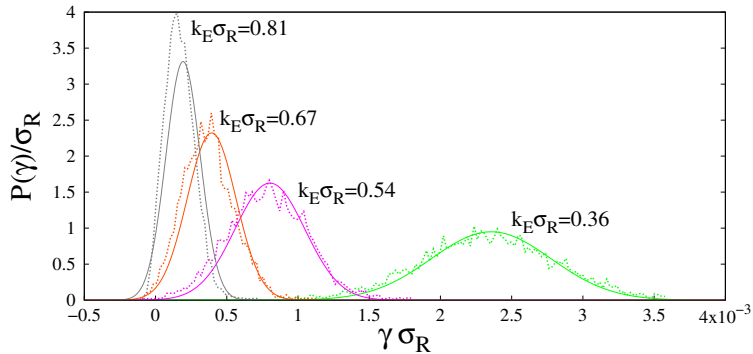}
\caption{
\label{fig-dist-simpleouv}
(Color online)
Probability distributions of the Lyapunov exponents obtained by transfer matrix calculations (dotted lines)
with the disorder parameters of Fig.~\ref{fig-simpleouv}, $V_R>0$
(space step of $\Delta z=0.1 \sigmar$ and total system size of $\Ltot=13333 \sigmar$
with random initial conditions), and various energies (indicated on the figure).
The solid lines are the corresponding theoretical distributions~(\ref{eq:distrib}).
}
\end{figure}
Figure~\ref{fig-simpleouv} shows the average values of the Lyapunov exponent, which we write $\av{\lyap}(E)$ in this paragraph.
For the sake of completeness, we show in Fig.~\ref{fig-dist-simpleouv}
some distributions of the Lyapunov exponents $\lyap(E)$ obtained for the different realizations of the potential at four values of the energy, together with the Gaussian distribution
\be
\label{eq:distrib}
P(\lyap)=\frac{1}{\sqrt{2\pi} \Delta_{\lyap}} \exp\left[-\frac{(\lyap-\av{\lyap})^2}{2\Delta_{\lyap}^2}\right]
\ee
with $\Delta_{\lyap}=\sqrt{\av{\lyap}/\Ltot}$ and $\Ltot$ the system size,
which is expected for $\delta$-correlated disorder when $\Ltot \gg 1/\av{\lyap}$.
For low energy, we find a good agreement.
Since $\lyap(E)=-\ln(T)/2\Ltot$,
where $T$ is the transmission probability of the wave through a sample
of finite length $\Ltot$,
it shows that
the transmission probability $T$ follows
a log-normal distribution also in the correlated disorder we are considering.
At high energy ($\kE \sigmar \gtrsim 0.8$) the numerical distributions differ from
Eq.~(\ref{eq:distrib}).
This is expected because for $\kE \sigmar =0.81$,
we find $1/\av{\lyap}\sim 5000 \sigmar$, which is of the order of magnitude of
$\Ltot=13333 \sigmar$.

\begin{figure}[t]
\center
\sidecaption
\includegraphics[width=0.75\textwidth]{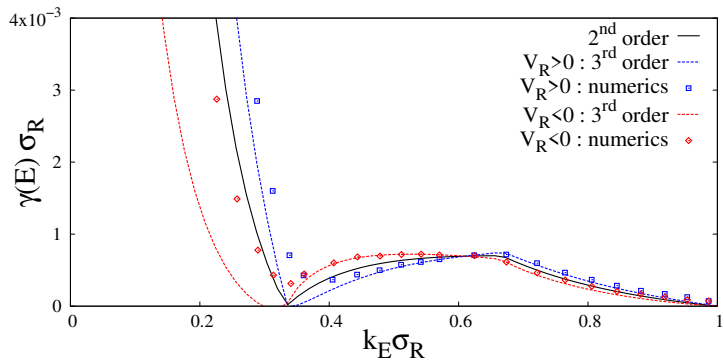}
\caption{
\label{fig-dblouv}
(Color online)
Lyapunov exponent for
speckles with the tailored correlation function~(\ref{eq:C2-2ouv}) with $\rho=1/3$,
using the same methods and parameters as in Fig.~\ref{fig-simpleouv}.
Blue squares: numerics with $\Vr>0$;
Red diamonds: numerics with $\Vr<0$;
Solid black line: order $2$ in the phase formalism;
Dotted blue and red lines: order $3$ in the phase formalism.
}
\end{figure}

As discussed above, for standard disorder, \ie\ with a power spectrum $\TFCor{2}$
that is a constant or decreasing function of $k$,
the Lyapunov exponent $\lyap^{(2)}(E)$ decreases with the energy $E$.
Let us now consider the double-slit configuration, of
correlation function given by Eq.~(\ref{eq:C2-2ouv}).
Inserting the latter into Eq.~(\ref{eq:lyap-ordre2}),
we find that, for $\rap>0.25$, $\lyap^{(2)}(E)$ shows an increase in a certain
interval of $k$, which is all the more pronounced that $\rho$ approaches $1$.
This indicates that in those tailored potentials, $\lyapE{E}$
can counter-intuitively increase with energy.
In order to study this effect precisely, we performed numerical (transfer-matrix approach) and analytical (phase formalism approach) calculations for the considered tailored speckle potential
with $\rap=1/3$, as done for the standard speckle potential.
The results of the numerical calculations (blue squares and red diamonds) and of analytical calculations up to order three in the phase formalism (dotted blue and red lines) are shown on Fig.~\ref{fig-dblouv}. They confirm that, for both blue and red detunings, $\lyapE{E}$ exhibits an increase with $E$ for $\kE \sigmar \in [0.35,0.6]$.
As for the standard speckle potential, the numerical results follow the trend of the $2^{\textrm{nd}}$ order term in the phase formalism (solid black line in Fig.~\ref{fig-dblouv}).
The $3^{\textrm{rd}}$ order term is given in the appendix [Eq.~(\ref{eq:o3-dblslit})].
For $\kE \sigmar \in [0.4,1]$, it accounts very well for the discrepancy between the numerics and the $2^{\textrm{nd}}$ order term (solid black line).
For low energy (\ie\ $\kE \sigmar \lesssim 0.4$), the $4^{\textrm{th}}$ and higher order terms play a more important role, which is expected as we are approaching the limits of validity of the perturbative development.

In the above double-slit configuration,
$\lyapE{E}$ has a slope break near its maximum [see Eqs.~(\ref{eq:C2-2ouv}), and (\ref{eq:o3-dblslit}) and Fig.~\ref{fig-dblouv}], which is reminiscent of the sharp edges of the pupil function. As it will presumably be inconvenient for experimental observations,
\begin{figure}[t]
\center
\sidecaption
\includegraphics[width=0.75\textwidth]{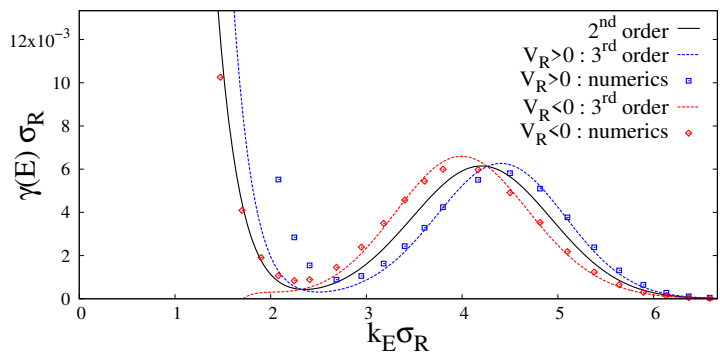}
\caption{
\label{fig-dblgauss}
(Color online)
Lyapunov exponent for
speckles with the tailored correlation function~(\ref{eq:C2-gauss}) with $\kappaNot= 8.88$ and $\Vr=\pm 0.72(\hbar^2/m \sigmar^2)$,
using the same methods as in Figs.~\ref{fig-simpleouv} and \ref{fig-dblouv}
(with a total system size of $\Ltot=30880 \sigmar$ and random initial conditions) averaged over 500 disorder realizations.
Blue squares: numerics with $\Vr>0$;
Red diamonds: numerics with $\Vr<0$;
Solid black line: order $2$ in the phase formalism;
Dotted blue and red lines: order $3$ in the phase formalism.
}
\end{figure}
we now consider the double-Gaussian configuration, which is obtained using
two mutually coherent Gaussian beams shone onto an infinite diffusive plate,
giving the power spectrum~(\ref{eq:C2-gauss})~\cite{piraud2012b}.
For this configuration, $\lyap^{(2)}(E)$ shows an increase when $\kappaNot \gtrsim 5.3$, which is all the more marked than $\kappaNot$ is large.
In Fig.~\ref{fig-dblgauss}, we show the Lyapunov exponents obtained in this case for $\kappaNot= 8.88$, with transfer matrices (blue squares and red diamonds) and with the phase formalism, up to order $2$ (solid black line) and up to order $3$ (dotted blue and red lines;
see Eq.~(\ref{eq:o3-dblgauss}) in the appendix for the $3^{\textrm{rd}}$ order term).
In this configuration, we recover the same trend as in the other configuration, namely
the Lyapunov exponent shows a significant increase in a given energy window,
$\kE \sigmar \in [2.3,4.2]$, 
the second order term, $\lyap^{(2)}(E)$, captures the main physics,
and the discrepancy between the numerical results and $\lyap^{(2)}(E)$ are well accounted
by the third order term, except at very low energy where the perturbative expansion breaks down.
As expected, the behavior of $\lyap(E)$ is smoother for the double-Gaussian configuration compared to the double-slit configuration.

\section{Observation scheme with ultracold atoms}
In order to probe the nonmonotonous behavior of $\lyap (E)$ discussed above,
one can use 
ultracold atoms, which proved a good means to observe 1D AL of matter waves with pseudo-periodic~\cite{roati2008} and speckle~\cite{billy2008} potentials.
The preceding calculations of the Lyapunov exponent (a self-averaging quantity) directly apply to a 1D transmission scheme of a wave with fixed energy $E$.
In ultracold-atom experiments~\cite{billy2008,roati2008,kondov2011,jendrzejewski2012} however,
a matter-wavepacket
with a broad energy distribution should be considered, and
the measured quantity is
the density profile obtained after releasing the atoms in the disorder, which is not directly related to the above calculations.
The average stationary density of a noninteracting atomic gas, with initial negligible width, after evolution in the disorder reads~\cite{lsp2007,lsp2008,piraud2011}
$n_\infty(z) = \int dE\ \distE (E) P_\infty(z | E)$,
where $\distE (E)$ is the energy distribution of the atoms and
\begin{eqnarray}
P_\infty(z \vert E)
& = &\frac{\pi^2 \lyap}{8} \int_0^\infty\! \textrm{d}u\ 
u\ \sh (\pi u) \left[ \frac{1+u^2}{1+\ch (\pi u)} \right]^2
\nonumber \\
& & \times \exp\{- (1+u^2) \lyap |z| / 2 \}\,,
\label{eq:gogolin}
\end{eqnarray}
with $\lyap = \lyap^{(2)}(E)$ given by Eqs.~(\ref{eq:dvpt-ordres}) and (\ref{eq:lyap-ordre2}),
is the probability of quantum diffusion calculated in the weak disorder approximation~\cite{berezinskii1974,gogolin1976a,gogolin1976b}.

\begin{figure}[t]
\center
\sidecaption
\includegraphics[width=0.75\textwidth]{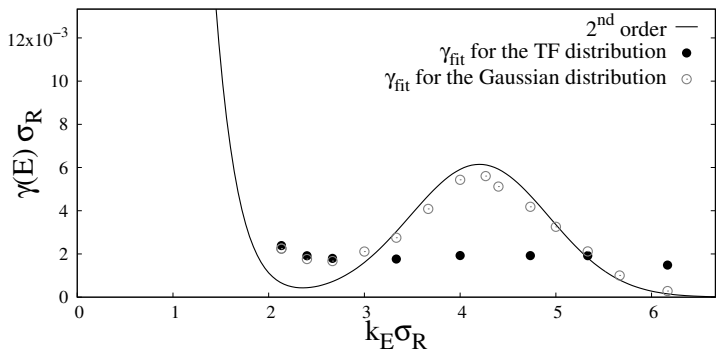}
\caption{
\label{fig:num-stat-TF}
(Color online)
Lyapunov exponent versus particle energy
for speckles
in the double-Gaussian configuration
with $\kappaNot=8.88$ and
$\Vr = \pm 0.72(\hbar^2/m\sigmar^2)$,
obtained from Eq.~(\ref{eq:lyap-ordre2}) (solid black line)
and from fits of Eq.~(\ref{eq:gogolin}) to the numerical data (points).
The figure shows the values of $\lyapfit$ extracted from the average profile $\av{n_\infty(z)}$,
with initial
  Thomas-Fermi [TF, Eq.~(\ref{eq:distTF}), full dots]
  and
  Gaussian  [Eq.~(\ref{eq:wigngauss}), open circles]
momentum distributions.
}
\end{figure}

A first attempt to observe the nonmonotonous behavior of $\lyap(E)$ may be to
consider the experimental scheme of Ref.~\cite{billy2008}.
In this case an interacting condensate is first produced in the Thomas-Fermi regime in a harmonic trap of frequency $\omega$, and the trap is then switched-off at time $t=0$.
In a first stage the expansion of the atoms is driven by their interaction energy,
and one can neglect the disordered potential.
For $t \gg 1/\omega$, it produces an almost noninteracting gas
with momentum distribution~\cite{kagan1996,castin1996}
\be
\mathcal{D}_{\textrm{i}} (p) = ({3}/{4\pmax}) \ftheta{1 - ( p / \pmax)^2 }.
\label{eq:distTF}
\ee
We have performed numerical integration of the time-dependent Schr\"odinger equation for a particle with the initial momentum distribution~(\ref{eq:distTF}) in the two-Gaussian tailored disordered potential with correlation function~(\ref{eq:C2-gauss}) and disorder parameters as in Fig.~\ref{fig-dblgauss}, for six realizations of the disordered potential [three with blue ($\Vr>0$) and three with red ($\Vr<0$) detuning]\footnotemark\footnotetext{As in Refs.~\cite{piraud2011,piraud2012b}, 
we use a Crank-Nicolson algorithm with the following numerical parameters:
  space step $\Delta x = 0.03\sigmar$,
  time step $\Delta t = 1.1\hbar/E_{\sigmar}$,
  boxes of size $12\times 10^3 \sigmar$.}.
After averaging the stationary density profiles over the six realizations, we fit $\ln[P_\infty(z)]$ as given by Eq.~(\ref{eq:gogolin}) to $\ln[\av{n_\infty}(z)]$ with $\lyap$ as the only fitting parameter\footnotemark\footnotetext{The fits are performed in the space windows $-300\sigmar < z < -50\sigmar$ and $+50\sigmar < z < +300\sigmar$, corresponding to an experimentally accessible width of 1 mm for $\sigmar= 1.6\, \mu m$.}.
The results, plotted on Fig.~\ref{fig:num-stat-TF}, show that the fitted Lyapunov exponent (black dots) slightly decreases with $\kmax$
and saturates roughly beyond the minimum of the calculated $\lyap^{(2)}$.
This is because the long distance behavior of $n_\infty(z)$ is dominated by the energy components with the largest localization lengths, \ie\ those with the smallest $\lyap (E)$~\cite{lsp2007,lsp2008,piraud2011}.
Therefore this scheme does not
enable us to probe the region where $\lyap (E)$ increases.

In order to observe the upturn of $\lyapE{E}$, one can use an atomic energy distribution much narrower in energy and strongly peaked at a given $\Emaxgas$,
so that $n_\infty (z) \simeq P_\infty(z \vert \Emaxgas)$.
As discussed in Ref.~\cite{piraud2012b}, it can be realized by
either giving a momentum kick
to a noninteracting initially trapped gas
or using an atom laser,
both with a narrow energy width~\cite{robins2006,guerin2006,bernard2011}.
Using the
momentum distribution
\begin{equation}
\distvi(p)=({1}/{\sqrt{2\pi} \largdist}) \exp\left[{-( p - \centredist )^2 / 2\largdist^2}\right].
\label{eq:wigngauss}
\end{equation}
with
parameters relevant to current experiments,
one can then extract the values of $\lyapE{\Emaxgas}$ by the same fitting procedure as above.
The results, displayed on
Fig.~\ref{fig:num-stat-TF} (grey circles, reproduced from Fig.~2(a) of Ref.~\cite{piraud2012b}),
show a strong increase of $\lyap(E)$ as a function of $\kat$,
which follows quite well Eq.~(\ref{eq:lyap-ordre2}).
The scheme, which requires a small change in current experiments hence allows one
to directly observe the nonmonotonic behavior of $\lyap(E)$ induced by the tailored correlations.

\section{Conclusion}
In summary, we have studied Anderson localization of noninteracting
ultracold matterwaves in correlated disordered potentials.
Tailoring of the disorder correlations
can strongly affect the localization behavior,
namely the localization length can increase with particle energy.
Here we focused on the 1D case, providing more details compared to earlier
works~\cite{plodzien2011,piraud2012b}.
In particular, we show that although the Born approximation reproduces the
overall behavior of the localization length, next-order terms are significant.
We have found good agreement between exact numerical calculations and perturbation theory up to order three.
We have compared two disorder configurations (`double-slit' and `double-Gaussian') to realize the above effect.
We finally discussed how to observe the effect with expanding ultracold atoms, explicitly showing that the scheme usually used in experiments~\cite{billy2008,roati2008,kondov2011,jendrzejewski2012}
should be adapted.

\begin{acknowledgement}
We thank A.~Aspect, Y.~Castin, P.~Chavel, D.~Cl\'ement, and J.-J.~Greffet
for discussions. 
This research was supported by
the European Research Council (FP7/2007-2013 Grant Agreement No.\ 256294),
Agence Nationale de la Recherche (ANR-08-blan-0016-01),
Minist\`ere de l'Enseignement Sup\'erieur et de la Recherche,
Triangle de la Physique
and
Institut Francilien de Recherche sur les Atomes Froids (IFRAF).
We acknowledge the use of the computing facility cluster GMPCS of the 
LUMAT federation (FR LUMAT 2764).
\end{acknowledgement}

\appendix
\paragraph{Third-order term in phase-formalism calculations}
For the double-slit configuration, the $f_3$ function that intervenes in Eq.~(\ref{eq:dvpt-ordres}) reads
\begin{eqnarray}
\fn{3}(\kappa) = \frac{\pi}{8(1-\rap)^3} &\Bigg\{ \Theta \left( \frac{1-\rap
}{2} - \kappa \right) f_{3,1}(\kappa) + \Theta \left( \frac{1-\rap
}{2} - \left| \kappa-\frac{\rap+1}{2} \right| \right) \nonumber\\
 &\times \bigg[ \Theta \bigg(\kappa - \frac{\rap+1}{2} \bigg) f_{3,2} (\kappa) + \Theta \left(  \frac{\rap+1}{2}-\kappa \right) f_{3,3} (\kappa) \bigg] \Bigg\},
\label{eq:o3-dblslit}
\end{eqnarray}
with
\begin{eqnarray*}
f_{3,1}(\kappa) &=& g(2)-g[2(1- \kappa)]-2g(2 \kappa) +2 g\left(1-\rap\right) -g\left(2\rap\right) -2 g\left(1-\rap-2 \kappa\right) \\
&&  +g\left(1-2\kappa+\rap\right)
+g\left[2 \left(\kappa+\rap\right)\right] -g\left(1+2\kappa+\rap\right),
\end{eqnarray*}
\begin{eqnarray*}
f_{3,2}(\kappa) &=& -g(2\kappa)-g\left(1-\rap\right)+g\left(2\rap\right)
+ g\left(2 \kappa+1-\rap\right)\\
&& + g\left(2 \kappa-2\rap\right) -g\left(1+\rap\right)+g \left(1-2 \kappa+\rap\right),
\end{eqnarray*}
and
\begin{eqnarray*}
f_{3,3}(\kappa) &=& g(2)-g[2(1- \kappa)]-g(2 \kappa)+g\left(1-\rap\right)-g\left(-1+2 \kappa-\rap\right)\\
&&-g\left(1+\rap\right)+g\left(2 \kappa-1-\rap\right)
\end{eqnarray*}
where $g(x)\equiv x\ln(x)$.

For the double-Gaussian configuration, we find~:
\begin{eqnarray}
\label{eq:o3-dblgauss}
\fn{3}(\kappa) &=& \frac{1}{32\sqrt{3}} \Bigg\{4\pi e^{-\frac{4}{3} \kappa^2} \Erfi{\frac{\kappa}{\sqrt{3}}}
+2\pi e^{- \kappa^2} \times\\
&&\Bigg[ \Big\{e^{-\frac{1}{3}(\kappa+\kappaNot)^2} \Erfi{\frac{\kappa+\kappaNot}{\sqrt{3}}}
+e^{-\frac{1}{3} (\kappa-\kappaNot)^2} \Erfi{\frac{\kappa-\kappaNot}{\sqrt{3}}} \Big\} \nonumber\\
&&+ 2e^{-\frac{\kappaNot^2}{4}} \cosh(\kappa \kappaNot)
\Big\{e^{-\frac{1}{3} (\kappa+\frac{\kappaNot}{2})^2} \Erfi{\frac{\kappa+\kappaNot/2}{\sqrt{3}}} \nonumber\\
&&
+ e^{-\frac{1}{3} (\kappa-\frac{\kappaNot}{2})^2} \Erfi{\frac{\kappa-\kappaNot/2}{\sqrt{3}}} \Big\} \Bigg]  \Bigg\} \nonumber
\end{eqnarray}
where $\Erfi{z} \equiv -i \, \Erf{iz}$ is the imaginary error function,
with $\Erf{x}\equiv 2 \int_0^x dt \, e^{-t^2}/\sqrt{\pi}$ the error function.

\end{document}